\documentclass[10pt,conference,final]{IEEEtran}

\usepackage[T1]{fontenc}
\usepackage{adjustbox}
\usepackage{pgfplots}
\usepackage{tikz}
\usepackage{algorithmic}
\usepackage{amsfonts}
\usepackage{amsmath,amssymb,amsfonts}
\usepackage{balance}
\usepackage{booktabs}
\usepackage{boxedminipage}
\usepackage{cite}
\usepackage{colortbl}
\usepackage{color}
\usepackage{filecontents}
\usepackage{float}
\usepackage{framed}
\usepackage{graphicx}
\usepackage{hyperref}
\usepackage{import}
\usepackage{listings}
\usepackage{multirow}
\usepackage{microtype}
\usepackage{paralist}
\usepackage{pgfplotstable}
\usepackage{pgfplots}
\usepackage{qtree}
\usepackage{relsize}
\usepackage{subfig}
\usepackage{tabularx}
\usepackage{array}
\usepackage{textcomp}
\usepackage{tikz}
\usepackage{url}
\usepackage{xcolor}
\usepackage{xr}
\usepackage{xspace}
\xspaceaddexceptions{\%}
\xspaceremoveexception{-}
\usepackage{fixltx2e}

\usetikzlibrary{fadings}
\usetikzlibrary{pgfplots.dateplot,positioning}

\xspaceaddexceptions{\%}
\xspaceremoveexception{-}
\def\BibTeX{{\rm B\kern-.05em{\sc i\kern-.025em b}\kern-.08em
    T\kern-.1667em\lower.7ex\hbox{E}\kern-.125emX}}

\definecolor{color0}{rgb}{0.886274509803922,0.290196078431373,0.2}
\definecolor{color1}{rgb}{0.203921568627451,0.541176470588235,0.741176470588235}
\definecolor{color2}{rgb}{0.556862745098039,0.729411764705882,0.258823529411765}
\definecolor{color3}{rgb}{0.00392156862745098,0.466666666666667,0.756862745098039}
\definecolor{color4}{rgb}{0.456862745098039,0.429411764705882,0.658823529411765}

\lstdefinestyle{diffs}{
  belowcaptionskip=1\baselineskip,
  breaklines=true,
  frame=L,
  xleftmargin=\parindent,
  showstringspaces=false,
  basicstyle=\footnotesize\ttfamily,
  moredelim=**[is][\it\color{red}]{@}{@},
  moredelim=**[is][\it\color{green}]{H}{H},
  identifierstyle=\color{blue},
  stringstyle=\color{orange},
  keywordstyle=\bfseries\color{green!40!black},
  commentstyle=\itshape\color{purple!40!black},
}



\newcommand{\DefineRQ}[2]{%
  \expandafter\newcommand\csname rmk-#1\endcsname{#2}%
}
\newcommand{\Remark}[1]{\csname rmk-#1\endcsname}

\DefineRQ{RQ1}{Effects on testing quantity}
\DefineRQ{RQ2}{Effects on testing quality}
\DefineRQ{RQ3}{Fault coupling}
\DefineRQ{RQ4}{Mutant redundancy}

\begin{document}

\title{Does mutation testing improve testing practices?}

\author{\IEEEauthorblockN{Goran~Petrovi{\'c}}
\IEEEauthorblockA{\textit{Google Switzerland, GmbH} \\
Z\"{u}rich, Switzerland \\
goranpetrovic@google.com}
\and
\IEEEauthorblockN{Marko~Ivankovi{\'c}}
\IEEEauthorblockA{\textit{Google Switzerland, GmbH} \\
Z\"{u}rich, Switzerland \\
markoi@google.com}
\and
\IEEEauthorblockN{Gordon Fraser}
\IEEEauthorblockA{\textit{University of Passau}\\
Passau, Germany \\
gordon.fraser@uni-passau.de}
\and
\IEEEauthorblockN{Ren{\'e} Just}
\IEEEauthorblockA{\textit{University of Washington}\\
Seattle, USA \\
rjust@cs.washington.edu}
}

\maketitle

\begin{abstract}
Various proxy metrics for test quality have been defined in order to guide developers when writing tests. Code coverage is particularly well established in practice, even though the question of how coverage relates to test quality is a matter of ongoing debate.
Mutation testing offers a promising alternative: Artificial defects can identify holes in a test suite, and thus provide concrete suggestions for additional tests.
Despite the obvious advantages of mutation testing, it is not yet well established in practice. Until recently, mutation testing tools and techniques simply did not scale to complex systems. Although they now do scale, a remaining obstacle is lack of evidence that writing tests for mutants actually improves test quality. 
In this paper we aim to fill this gap: By analyzing a large dataset of almost 15 million mutants, we investigate how these mutants influenced developers over time, and how these mutants relate to real faults.
Our analyses suggest that developers using mutation testing write more tests, and actively improve their test suites with high quality tests such that fewer mutants remain.
By analyzing a dataset of past fixes of real high-priority faults, our analyses further provide evidence that mutants are indeed coupled with real faults. In other words, had  mutation testing been used for the changes introducing the faults, it would have reported a live mutant that could have prevented the bug.
\end{abstract}


\newcommand{\hl}[2]{\relax}

\newcommand{\google}[1]{\hl{red}{#1}}
\newcommand{\oss}[1]{\hl{green}{#1}}

\newcommand{\todo}[1]{\hl{red}{#1}}
\newcommand{\bob}[1]{\hl{olive}{#1}}
\newcommand{\rene}[1]{\hl{blue}{#1}}
\newcommand{\paul}[1]{\hl{violet}{#1}}

\def\lhs{lhs\xspace}
\def\rhs{rhs\xspace}
\newcommand\expr[1]{\texttt{#1}}

\def\unprod{unproductive\xspace}
\def\Unprod{Unproductive\xspace}
\def\prod{productive\xspace}
\def\Prod{Productive\xspace}

\def\mutationDataset{mutation dataset\xspace}
\def\MutationDataset{Mutation dataset\xspace}

\def\unprodStartPercentage{80\%\xspace}
\def\unprodEndPercentage{10\%\xspace}

\def\CompanyX{Google\xspace}








\def\nPLs{4\xspace}

\def\ratioCLsForPLs{90\%\xspace}

\def\nCLs{1.9 million\xspace}

\def\nDevs{30,000+\xspace}

\def\nBugsTotal{1765\xspace}
\def\nBugsTooLarge{222\xspace}
\def\nBugsNonRepro{41\xspace}
\def\nBugsFinal{1502\xspace}
\def\nBugsCoupled{1043\xspace}
\def\nBugsNotCoupled{459\xspace}
\def\ratioBugsCoupled{70\%\xspace}
\def\ratioBugsNotCoupled{30\%\xspace}

\def\nMutantsFate{95544\xspace}
\def\nLinesFate{24769\xspace}

\newtheorem{Ra}{Research Answer}
\def\ogap{-5pt}
\def\igap{-5pt}
\newenvironment{concl} {\vspace*{\ogap}\begin{framed}\vspace*{\igap}}
{\vspace*{\igap}\end{framed}\vspace*{\ogap}}

\newcommand{\z}{\phantom{0}}
\newcommand{\p}{$\scriptsize{\pm}$}

\def\<#1>{\codeid{#1}}
\newcommand{\codeid}[1]{\ifmmode{\mbox{\small\ttfamily{#1}}}\else{\small\ttfamily #1}\fi}
\newcommand{\codeidsmall}[1]{\ifmmode{\mbox{\smaller\ttfamily{#1}}}\else{\smaller\ttfamily #1}\fi}
\def\|#1|{\code{#1}}
\newcommand{\code}[1]{\ifmmode{\mbox{\ttfamily{#1}}}\else{\ttfamily #1}\fi}

\def\dJ{Defects4J\xspace}
\def\evo{EvoSuite\xspace}
\def\randoop{Randoop\xspace}
\def\major{Major\xspace}

\newcommand{\etal}{et al.\xspace}

\def\ngram{$n$-gram\xspace}
\def\ngrams{$n$-grams\xspace}

\def\ror{ROR\xspace}
\def\cor{COR\xspace}
\def\aor{AOR\xspace}
\def\sor{SOR\xspace}
\def\std{STD\xspace}
\def\lvr{LVR\xspace}
\def\evr{EVR\xspace}
\def\oru{ORU\xspace}
\def\lor{LOR\xspace}

\def\equM{equivalent\xspace}
\def\domM{dominator\xspace}
\def\trvM{trivial\xspace}
\def\subM{subsumed\xspace}

\def\equMShort{equi.\xspace}
\def\domMShort{dom.\xspace}
\def\trvMShort{triv.\xspace}
\def\subMShort{sub.\xspace}

\newcommand\dtype[1]{\texttt{#1}\xspace}
\newcommand\feature[1]{\textsf{\small{#1}}\xspace}

\newcommand\op[1]{$\textnormal{Op}_#1$}
\newcommand\bop[1]{$\bm{\textnormal{\textbf{Op}}_#1$}}
\newcommand{\mypara}[1]{\vspace*{2pt}\noindent\textbf{#1}\quad}

\newcommand\colH[1]{\multicolumn{1}{c}{\textbf{#1}}}
\newcommand\colsH[1]{\multicolumn{2}{c}{\textbf{#1}}}

\def\rowSmallSample{\rowcolor{lightgray}}
\def\rowLargeSample{\relax}

\newcolumntype{H}{>{\setbox0=\hbox\bgroup}c<{\egroup}@{}}

\def\total{Total}
\def\retained{Retained}

\newcommand\mop[2]{\texttt{#1} $\longmapsto$ \texttt{#2}}
\newcommand\context[1]{\|#1|}

\newenvironment{result}%
{\medskip
	\noindent
	\let\emph=\textbf
	\begin{boxedminipage}{\columnwidth}\begin{center}\em}%
		{\end{center}\end{boxedminipage}%
	\medskip
}

\section{Introduction}

Testing is an essential part of software development, and software developers need guidance in how much testing they need to do, and where to add more tests. Various proxy metrics for test suite quality have been defined over time with the aim to provide this guidance. Out of these, code coverage is probably best established in practice: Executing code is a prerequisite for revealing faults, code coverage can be easily visualized, for example by highlighting covered lines in a different color than uncovered lines, coverage is cheap to compute, and it is well supported by commercial-grade tools~\cite{IvankovicPJF2019}. 

There are, however, downsides to coverage: Quantifying code quality based on
code coverage alone leads to questionable estimates, whose general utility and
actionability are a matter of controversy in the research community~\cite{Inozemtseva2014,Gopinath2014,kochhar2015code}. Coverage-adequate test suites, which
satisfy all test goals, are not the norm nor should they be. In practice,
developers only satisfy a fraction of the test goals, but adequate thresholds
for code coverage ratios are inherently arbitrary and a matter of much
debate~\cite{CoverageGoogle,marick1999misuse}. Code coverage is also easily fooled, as it only determines whether code has been executed, regardless of how well its behavior has been checked.

These downsides are overcome by mutation testing~\cite{acree1979mutation,DeMillo:1978:HTD:1300736.1301357}: A test suite is evaluated on a set of systematically generated artificial faults ({mutants}). Any {surviving mutant} that is not detected by the test suite constitutes a concrete test goal, pointing out possible ways to improve the test suite. While the idea of mutation testing is appealing, adoption in practice has long been hampered by scalability issues: Even simple programs may result in large numbers of mutants. Performing traditional mutation testing, which evaluates all possible mutants, for a large code base is impracticable. For example, at \CompanyX, 500,000,000 tests are executed per day, gate-keeping 60,000 code changes. 
However, as a result of decades of research on mutation testing, it is now
possible to apply mutation testing even at such
scale~\cite{PetrovicI2018,PetrovicIFJ2021}.

While the computational challenges should no longer prevent adoption in practice, there remains uncertainty about whether the expected benefits of showing mutants to developers manifest:
For a large system many mutants can be produced, yet only few of these can be shown to developers. Do developers actually improve their test suites when shown these mutants, and do tests written for these mutants have the potential to help prevent real faults? In this paper, we aim to shed light on these questions of central importance.

Using a dataset of almost 15 million mutants, created in industrial software projects at \CompanyX over a duration of six years, we investigate the effects of showing these mutants as test goals to developers. In particular, we aim to answer the following research question:

\begin{itemize}

\item {\bf RQ1 \Remark{RQ1}}. How does continuous mutation testing
  affect how much test code developers produce?

\end{itemize}

Our data shows that developers working on projects exposed to mutation testing over a longer period of time tend to write more tests for their code (Section~\ref{sec:rq1}). This raises an immediate follow-up question: Are they writing \emph{good} tests?

\begin{itemize}

\item {\bf RQ2 \Remark{RQ2}}. How does continuous mutation testing
  affect the survivability of the mutants on a project?

\end{itemize}

Our data shows that the tests that developers add when exposed to mutants are effective: The more mutants they act on, the fewer mutants survive for new changes over time as the effectiveness of their test suites improves (Section~\ref{sec:rq2}); the additional tests written improve the test suite's ability to detect mutants beyond the ones for which they were written.

Naturally, this leads to the next question: Are tests that are effective at detecting mutants also effective at detecting real faults? Mutants are worth writing test cases for only if they are \emph{coupled} with real software faults, i.e., test suites that detect mutants would also detect real faults. The third research question is centered around this issue:

\begin{itemize}

\item {\bf RQ3 \Remark{RQ3}}. Are reported mutants coupled with real software
faults?
 Can tests written based on mutants improve test effectiveness for real
software faults?

\end{itemize}

Using a dataset of historical faults and fixes, our data shows that had mutation
testing been used for the fault-introducing changes, it would have reported a
live mutant that is killed by tests in the fault-fixing change
(Section~\ref{sec:change_report}). It is thus likely that mutants, had they been
reported, could have prevented the fault by guiding the developers to
investigate the mutants and write tests for them.

Finally, one of the reasons that modern mutation testing systems scale is that
they do not generate and analyze all possible mutants---only a small sample. In
particular, in our dataset developers were never shown more than one mutant per
line of code. This leads to the fourth research question:

\begin{itemize}

\item {\bf RQ4 \Remark{RQ4}}. Are the mutants generated for a given line redundant? Is
  it sufficient in practice to select a single mutant per line?
 
\end{itemize}

Our data shows that most mutants share a majority fate
(Section~\ref{sec:majority}). If a single mutant is killed in a line, most
likely all mutants in that line will be killed. Conversely, if a single mutant
survives in a line, most likely all mutants in that line will survive. This
backs the intuition and design choices behind Google's mutation testing system,
in particular the practice of generating only a single mutant per line.

Overall, our results, for the first time, provide strong evidence that mutants are coupled with real faults that matter in practice, and that showing mutants as test goals to developers leads to them writing more and better tests.

\section{Mutation Testing at Google}
\label{sec:background}

%
Traditionally, mutation testing is performed by generating all viable mutants for a software under test, and then running its test suite in an attempt to kill those mutants, i.e., to have at least one test that fails on the mutant code. The output of this process is the mutation score (i.e., the ratio of killed mutants), a higher-is-better measure.
Considering the scale of software systems at Google, this
traditional approach is computationally infeasible. Even if it were
feasible, the mutation score itself is difficult to act on by developers.
As a consequence, mutation testing at \CompanyX has evolved compared to the
traditional approach~\cite{PetrovicI2018}.

Google's mutation testing system currently supports 10 programming languages:
C++, Java, Go, Python, TypeScript, JavaScript, Dart, SQL, Common Lisp, and
Kotlin.  This section describes the main distinguishing features of the
mutation testing system, and we refer an interested reader to a full
description of the system for further details~\cite{PetrovicIFJ2021}.

The key difference at \CompanyX is that mutation testing is integrated into the
code review process: Developers send their code changes for code review
in form of \emph{changelists}. A changelist is subject to various
static and dynamic analyses that produce \emph{code findings}, which are reported to the
developer and the reviewers assigned to that changelist. An example analysis is
line coverage, which is reported at the level of the changed files as well as
the actual code changes. Mutation testing is integrated as another such finding: A
selection of mutants within the changed code of a changelist, which are not
killed by the existing tests, are shown as findings. These live mutants thereby
serve as concrete test goals, which a developer can satisfy by adding tests. 
To avoid cognitive
overload, no more than seven mutants are reported per file in a changelist.
Reviewers can mark findings as particularly important by clicking on the
corresponding `Please fix' link, and developers can provide feedback about
findings that are not actionable using a `Not useful' link. Developers can update
the changelist based on findings and reviewer comments.
Once the reviewers approve, a changelist is \emph{submitted} (merged) into the main
source tree.

Mutating only changed code is key to reducing the computational costs, but
further optimizations are necessary to make mutation testing feasible
and actionable.
First, mutants are only generated in lines that are covered by tests,
since lack of coverage is already reported by the coverage analysis.
Second, only one mutant is generated per line.
Third, in order to increase the
chances of generating productive mutants~\cite{PetrovicIKAJ2018},
suppression rules filter out code that cannot result in productive mutants
(e.g., logging statements). The developer feedback informs these heuristics for
mutant productivity.
Finally, after applying the suppression rules, an important and technically challenging aspect
is the choice of mutation operator to generate a mutant
for a given line. Our mutation system implements five mutation operators:
AOR (Arithmetic operator replacement),
LCR (Logical connector replacement),
ROR (Relational operator replacement),
UOI (Unary operator insertion), and
SBR (Statement block removal).
The mutation system selects a mutation operator for a given line
based on historical information on whether mutants in similar context
survived and whether developers thought they were productive. If a mutation
operator is not applicable to a line, another one is chosen, until one mutant
in the line is generated.
%
%

%

Many individual software projects are developed at \CompanyX, and each project can opt-in to use the mutation testing system.
When mutation testing is enabled for a project, each changelist in review for
that project will be analyzed for mutants, and live mutants will be reported
to the changelist author and reviewers. Over time, the same files may be
mutated repeatedly whenever they are part of a changelist under review. The
mutation system does not offer manual invocations or project-level
analyses---only the change-based approach.
Hence, developers only observe live mutants that are reported during code review.

Mutation testing has been deployed at \CompanyX for more than six years, and
some projects have been rolled in since the beginning, while many joined over
time. There have now been more than 15 million mutants generated, and hundreds
of thousands of analyzed changelists, providing a suitable dataset for analyzing
the long-term effects of mutation testing on the developers working on a project
employing it.

\section{Long Term Effects of Mutation Testing}
\label{sec:longtermeffects}

A core assumption of mutation testing is that mutants are actionable
test goals, and that reporting them has the desired effect on
developers:
They augment their test suites, and generally improve their testing
practices over time. 
%
%
We therefore wished to study whether developers exposed to mutants write more tests,
%
and whether these tests improve the test suite quality beyond simply covering additional code.
%
Thus, in this section we focus on the effects that mutants have on the tests
developers \mbox{write, and aim to answer the first two research questions:}

\begin{itemize}

\item {\bf RQ1 \Remark{RQ1}}. How does continuous mutation testing affect how
much test code developers produce?

\item {\bf RQ2 \Remark{RQ2}}. How does continuous mutation testing
  affect the survivability of the mutants on a project?

\end{itemize}

To answer these research questions, we performed a longitudinal
study that fully integrates mutation-based testing into the software-development
process.  Our study employs a direct intervention to form a treatment group
(developers who act on mutants) and uses a control group (developers who use
only code coverage).



\subsection{Datasets}

Although we are interested in how mutation testing affects developers, focusing
on individual developers for analysis is difficult: Developers tend to
multi-task, either working on a side project, maintaining an older code base, or still partially working on their previous engagement. 
Different codebases tend to differ greatly in their testing quality, especially when comparing side projects, older projects, and new engagements. Thus, instead of considering the behavior of individual developers we consider individual \emph{files} of source code. We created two datasets of files: One dataset based on files subjected to mutation testing and, as baseline, one dataset based on files subjected to only code coverage analysis.

\mypara{Mutant dataset} Google has used mutation testing since 2014. At the time of this writing, a total of 14,730,562 mutants have been generated for 662,584 code changes in 446,604 files. Our mutant dataset contains all of them.
For each mutant, the dataset contains information about the mutation operator, the programming language of the mutated file, the code location, and the results of mutation testing (i.e., whether the mutant was killed and whether a reviewer requested a test for it).

\mypara{Coverage dataset} As a baseline for comparison, we randomly sampled code changes from projects that did not use mutation testing, but had line coverage calculation enabled, within the same period of time. In total, the sample resulted in 8,788,791 code changes with 3,398,085 files.

\subsection{Methodology}

\subsubsection{RQ1}

We are interested in understanding the effects of mutants on tests written for
the mutated code. Therefore, for each file we need to quantify the amount of
testing related to that file. For each code change, we identified the edited
test files that correspond to that change, using the build system.
However, the raw number of edited lines of test code is not a reliable measure
because of differing programming languages, testing styles, categories of tests
and the specifics of testing and mocking frameworks. For example, in a
table-based testing approach, a test case may only consist of a single line of
code, whereas an equivalent test case, possibly including harness code, can
easily span tens or hundreds of lines.
Therefore, we quantify the amount of testing using the number of
\emph{test hunks}---the changed code hunks in the edited test files.
Hunks are groups of differing lines (edited or added) that intersperse
sequences of lines in common.

In addition, we measure the \emph{exposure} for each file in a code change,
computing the number of times the file has been changed \emph{and} had a code finding
reported before the change under analysis. Code findings are reported mutants
for the mutant dataset and code coverage results for the coverage dataset.
For example, if a change edits a file for the 10\textsuperscript{th} time, but
in the previous 9 times mutants were reported only in 3
instances, the \emph{exposure} would be 3 for that change.
This measure captures the exposure of developers to mutants: The
more often mutants had been reported in a file, the more exposure the
authors and reviewers developing that file collectively had to mutation testing.

In order to answer RQ1, we evaluate the number of test hunks for each file over
the file's exposure. If developers get accustomed to mutation testing and act on
mutants to improve the test suite, we expect the number of test hunks to be
larger for mutation than for the baseline. Further, we expect a positive
relationship between exposure and number of test hunks. Since we do not expect
that relationship to be linear and to show some saturation, we rely on rank
statistics. Specifically, we quantify the effect size using Spearman's rank
correlation coefficient $r_s$.
In addition, we compute the number of test hunks for individual changelists,
considering two timestamps: (1) when code findings are initially reported to the
developer during code review and (2) when the changelist is approved and
submitted. If code findings are actionable, we expect an increase in the number
of test hunks, after reporting the findings.

\subsubsection{RQ2}

RQ1 is concerned with testing quantity, but not quality.
Writing more tests does not necessarily improve the quality of a test suite.
For example, tests written simply for the sake of increasing code coverage tend
to be ineffective at detecting faults~\cite{marick1999misuse}.
In particular, if mutation testing generally improves test suite quality,
then we expect to observe several effects: First, reported
mutants should be killed by additional tests while a changelist is in code
review; second, over time, the ratio of mutants (generated for a new changelist)
that are already killed by the existing tests should increase; third, fewer mutants
should see `Please fix' requests by the reviewers.

As a proxy measurement of test quality, we use mutant \emph{survivability},
i.e., the probability of a generated mutant being killed. The mutant
survivability for a file in a code change is calculated as the ratio of
surviving mutants to the total number of generated mutants for that file.
A lower mutant survivability indicates a higher test suite quality.
We compute the mutant survivability for each file in the mutant dataset over
the \emph{exposure} of the mutated file. If the tests added due to mutation
testing lead to an improvement of the test suite quality, then we expect a
negative correlation between exposure and survivability, i.e., the tests should
become more effective at killing mutants over time---even for mutants that are
generated later for newly added or changed code. If tests are strictly added to
kill reported mutants, we expect no correlation because survivability should
remain fairly constant.

We further compare the number of live mutants first
reported to the developer with the final number of live mutants when the
changelist is submitted. If mutants are actionable test goals, we expect
that the number of live mutants at submission time is lower than the number of
originally reported mutants as a result of the additional tests added by the
developer.

In the code review system, reviewers can request developers to write tests for
mutants they consider important. If developers improve their test suites over
time, we would expect the ratio of such requests (for all reported mutants) to
gradually decrease. 
Therefore, we also compute the ratio of requested fixes for mutants over
exposure.

\subsection{RQ1: \Remark{RQ1}}

\label{sec:rq1}

\begin{figure}
	\centering 	\includegraphics[trim=0 0 0 30, clip, width=\columnwidth]{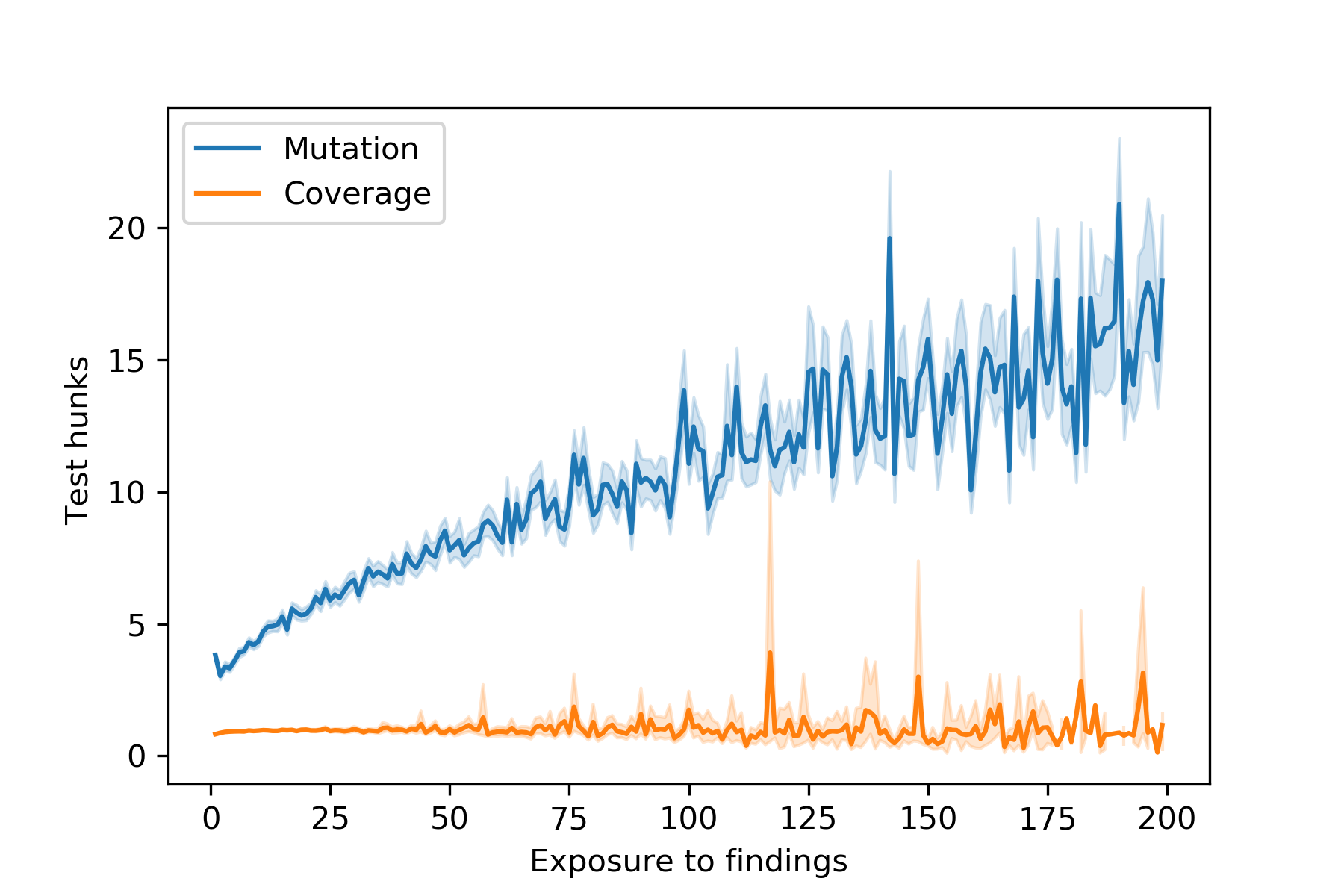}
	\caption {Test hunks changed per file as exposure to
  reported mutants vs. line coverage increases.}
	\label{fig:hunks}

    \vspace*{10pt}
	\centering 	\includegraphics[trim=0 0 0 20, clip, width=\columnwidth]{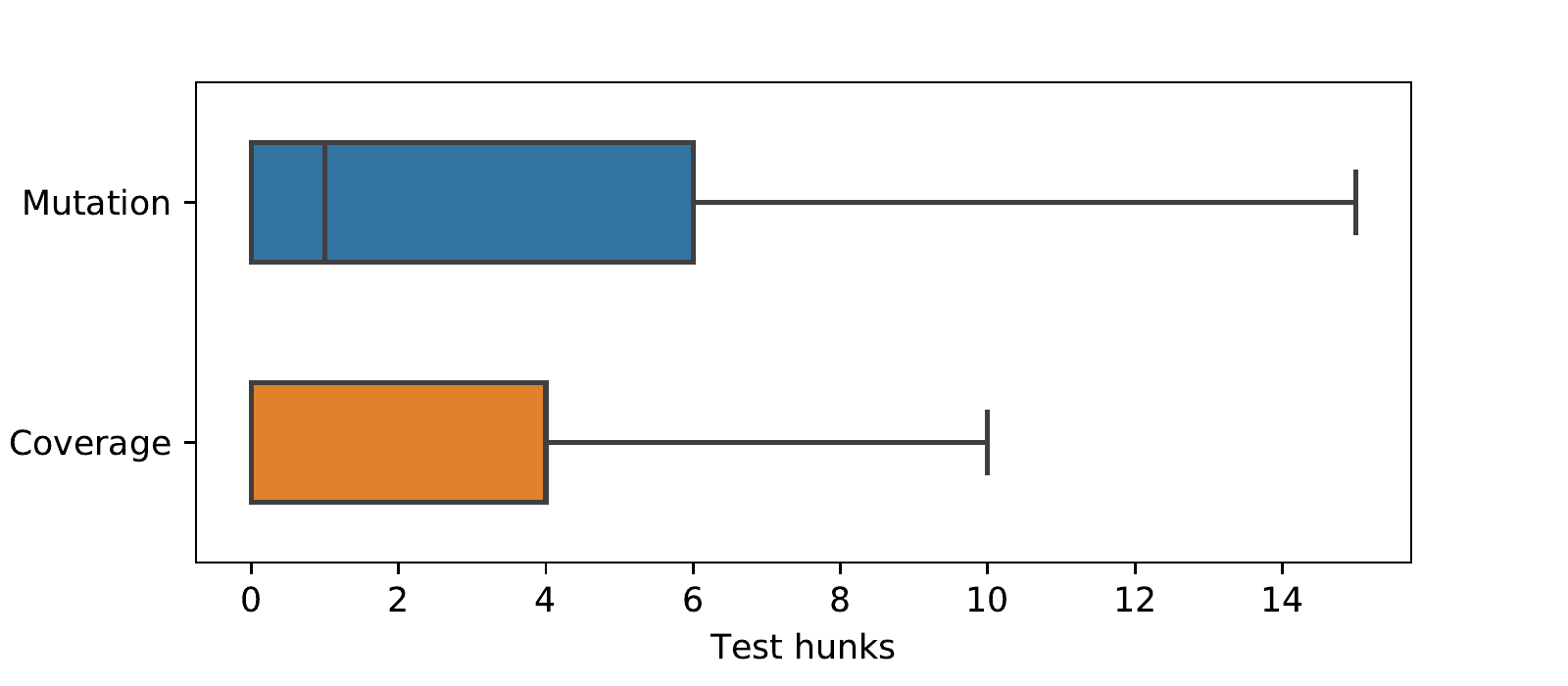}
	\caption {Test hunks changed per changelist after initially reporting
              mutants vs. line coverage. The median is 1 for mutation and 0 for
              coverage. Outliers (12\% outside of $\pm 1.5$ times the
              interquartile range) are not shown for clarity.}
	\label{fig:hunks_intervention}
\end{figure}

Figure~\ref{fig:hunks} shows the number of changed hunks of test code,
normalized by the number of files in the change, over exposure to code findings.
This figure shows the estimated central tendency (mean) and corresponding
confidence interval.
Note that the confidence intervals get wider for higher levels of exposure
for mutation because the number of files with such high levels
in the mutant dataset decreases. This means that accurately fitting a
trend for higher levels of exposure and identifying saturation is challenging.
However, there is a clear signal for the mutant
dataset: (1) The number of test hunks is larger than that for coverage.
(2) The larger the exposure, the more test hunks are changed on average.
Spearman's rank correlation between exposure and the
average number of changed test hunks is $r_s=0.9$ ($p< .001$).
In contrast, Figure~\ref{fig:hunks} shows no signal for the coverage
dataset. In fact, there is a weak negative correlation between exposure
and the average number of changed test hunks ($r_s=-0.24$, $p< .001$).
These results suggest that mutation testing has a positive long-term effect on
testing quantity.

Figure~\ref{fig:hunks_intervention} shows the number of test hunks changed in the
changelist between the initial state (when a developer sends it for review) and
the final state (when it passes the code review and is submitted). The median
number of changed test hunks for mutation is $1$, while for coverage it is $0$.
The difference between mutation and coverage is statistically
significant (Wilcoxon Rank Sum Test, $p<
.001$), and remains significant when normalized for changelist size.
This suggests that mutants are actionable and guide developers in writing
additional tests.

\begin{result}
  \textbf{RQ1:} As exposure to mutation testing increases,
                developers tend to write more tests.
\end{result}

\mypara{Alternative Hypotheses}
While our study uses an interventional design and a substantial dataset, it is
not a fully randomized controlled experiment. Hence, we account for
possible confounding factors, which might also explain the increase in testing
quantity. Specifically, we explored and rejected three alternative hypotheses.

\begin{enumerate}
\item\textit{Dispersed coverage dataset:}
The \emph{mutant} dataset contains all changes that were subject to mutation
testing, whereas the \emph{coverage} dataset contains only a sample of
changes that were subject to code coverage analysis. As a result, the coverage
dataset may miss individual changes to individual files. We downsampled the
mutant dataset to make it as dispersed as the coverage one and repeated our
analysis. The results led to the same conclusions for RQ1: There is a strong
positive relationship ($r_s=.82$, $p<0.001$) between exposure and testing
quantity, increasing our confidence that the
trend for the mutant dataset is indeed linked to mutation testing and that the
absence of that trend in the \mbox{coverage dataset is not an artifact of random
sampling.}

\begin{figure}
  \includegraphics[trim=0 0 0 30, clip, width=\columnwidth]{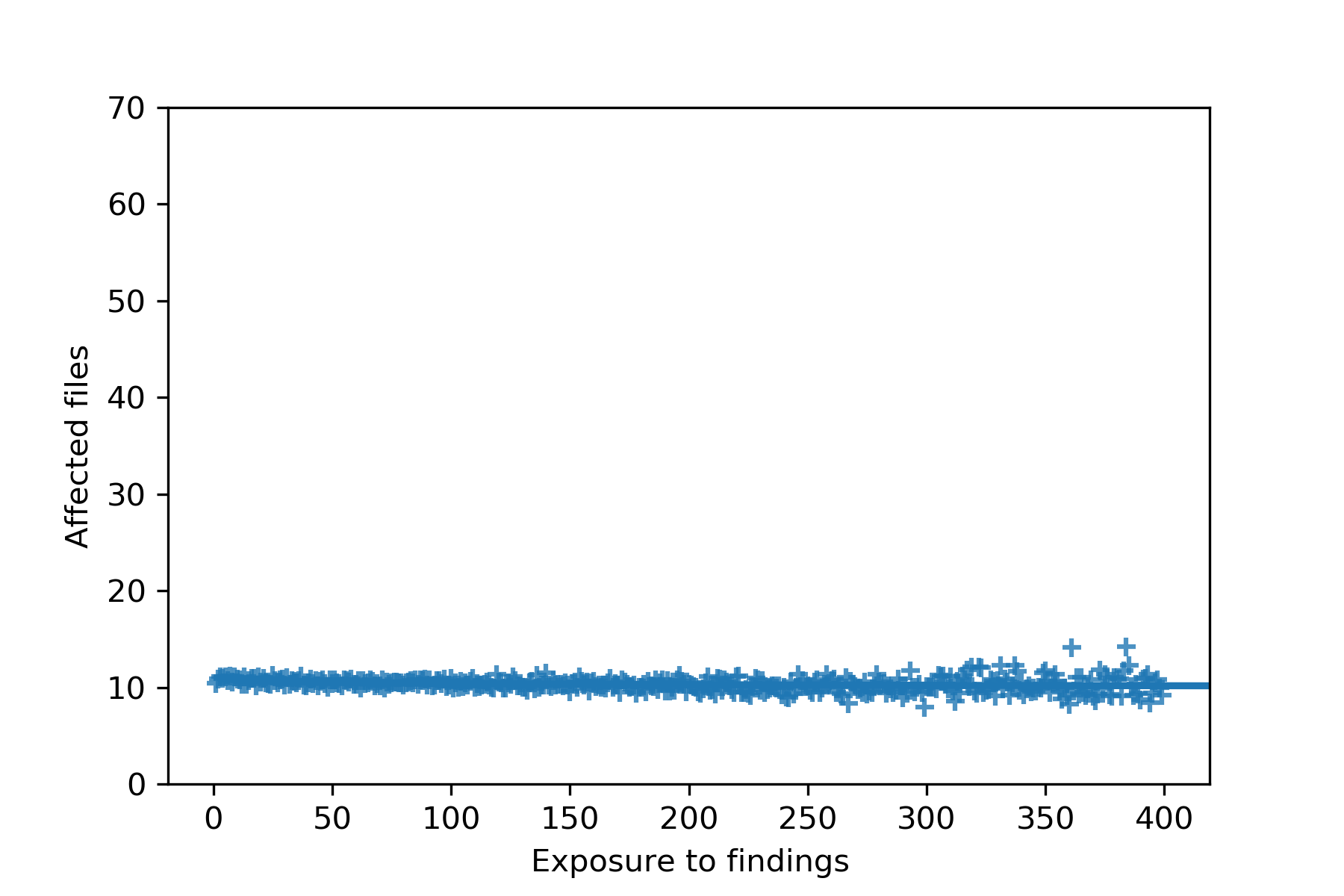}
  \caption {Files changed as exposure to mutants increases.}
  \label{fig:changes_mutation}
  \vspace*{10pt}
  \includegraphics[trim=0 0 0 30, clip, width=\columnwidth]{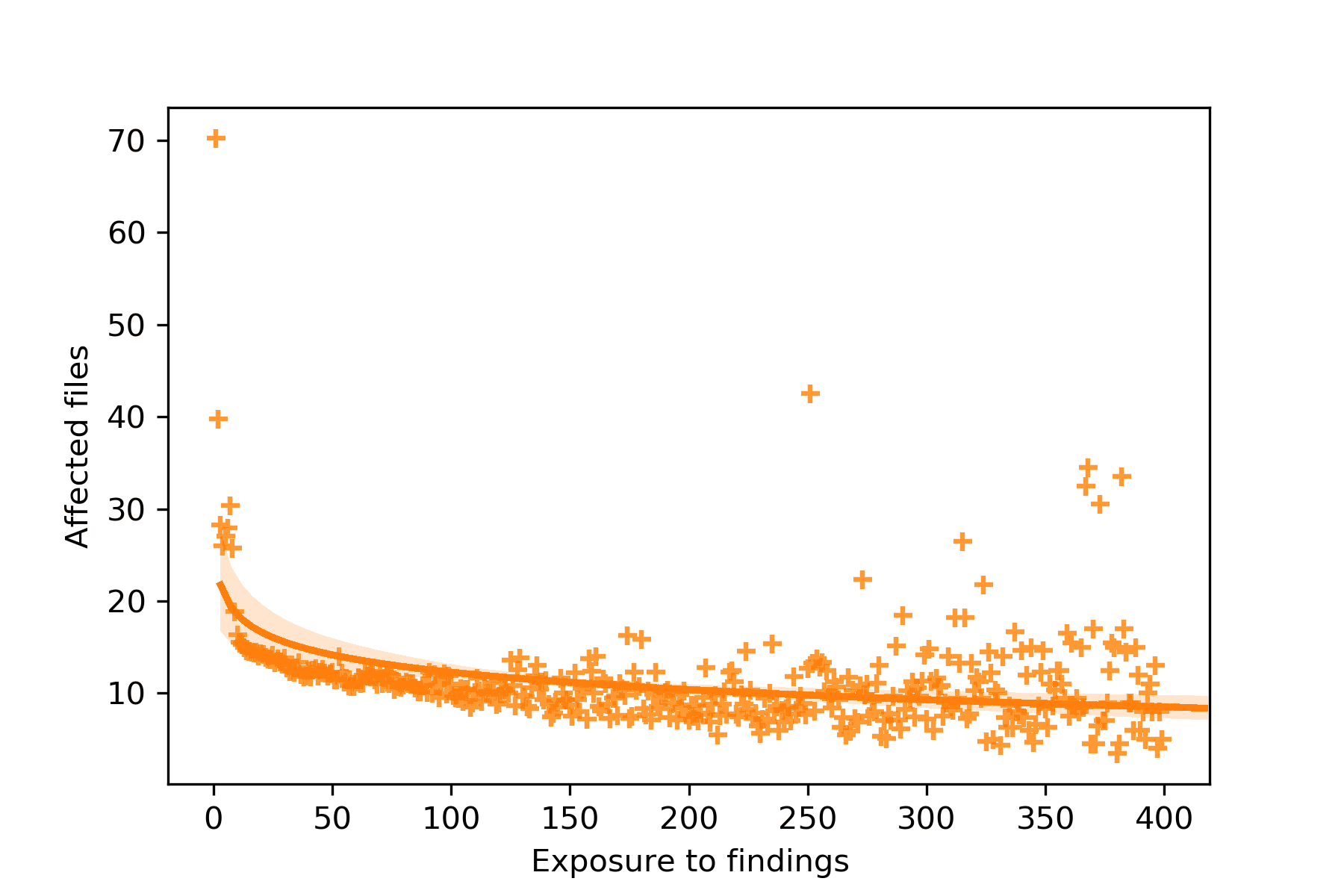}
  \caption {Files changed as exposure to line coverage increases.}
  \label{fig:changes_coverage}
\end{figure}

\item
\textit{Changelist size:}
Coverage analysis can be done on virtually any code change, while mutation
testing has some preconditions. In order for a mutant to be reported for a code
change, that change must contain covered source code, and the covered source
code must contain code elements for which mutants are not suppressed by the
mutation system's suppression rules.  It is thus expected that on average the
mutant dataset contains larger changelists than the coverage dataset---largely due to
differences in the lower tails.  If additional test hunks were just the
by-product of larger changelists, then we would expect normalizing for the
changelist size to uncover this fact, which it did not (neither on a
per-file nor on a per-changelist basis).

Changes to individual files are part of code changes of varying sizes. To ensure
that any effects observed are not a result of differences in the number of
changed files, Figure~\ref{fig:changes_mutation} and
Figure~\ref{fig:changes_coverage} show the number of files changed by a
changelist that contains an exposed files, as a function of exposure.  The plots
suggest that the numbers of affected files is not positively correlated with
exposure for the mutant dataset; indeed the number of affected files tends to
slightly decrease over time for both datasets ($r_s=-0.26$ for the \emph{mutant}
dataset and $r_s=-0.34$ for the \emph{coverage} dataset).

Overall, the results show that it is very unlikely that the changelist size as a
function of exposure confounds the observed effects.

\item
\textit{Tests for code coverage:}
Developers who received mutant information also received coverage information,
which could confound our results. In order to determine
whether developers wrote tests for the reported mutants or simply to
increase code coverage for their changelists, we analyzed the differences in
code coverage for the changed code between the initial review state and the
final submit state and correlated it with exposure. The results show that
code coverage is largely stable over time: For the \emph{coverage} dataset, there is a
negligible correlation with exposure ($r_s=-0.02$, $p<0.001$); for the
\emph{mutant} dataset, there is a weak negative correlation
($r_s=-0.17$, $p<0.001$).
%
%
%
This makes it unlikely that developers exposed to mutants wrote tests to increase code coverage.
Rather, they wrote tests to kill the reported mutants.

\end{enumerate}

\subsection{RQ2: \Remark{RQ2}}

\label{sec:rq2}

\begin{figure}[t]
  \includegraphics[trim=0 0 0 20, clip, width=\columnwidth]{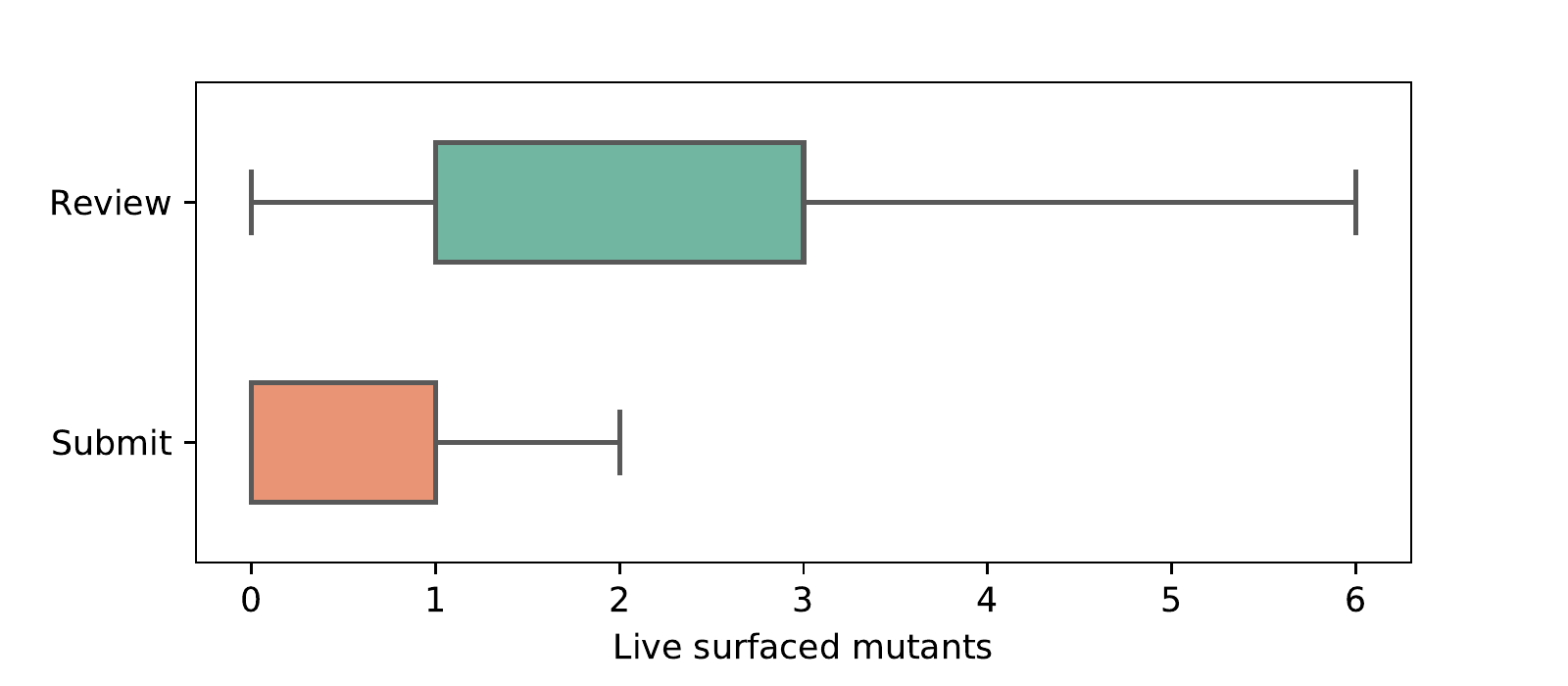}
  \caption {Number of surfaced mutants that are live at the beginning of the
            code review (\emph{Review}) and at the end of the code review (\emph{Submit}).
            The median is 1 for Review and 0 for Submit.
            Outliers (11\% outside of $\pm 1.5$ times the interquartile range)
            are not shown for clarity.}
  \label{fig:killed}
\end{figure}

\begin{figure}[t]
  \includegraphics[trim=0 0 0 30, clip, width=\columnwidth]{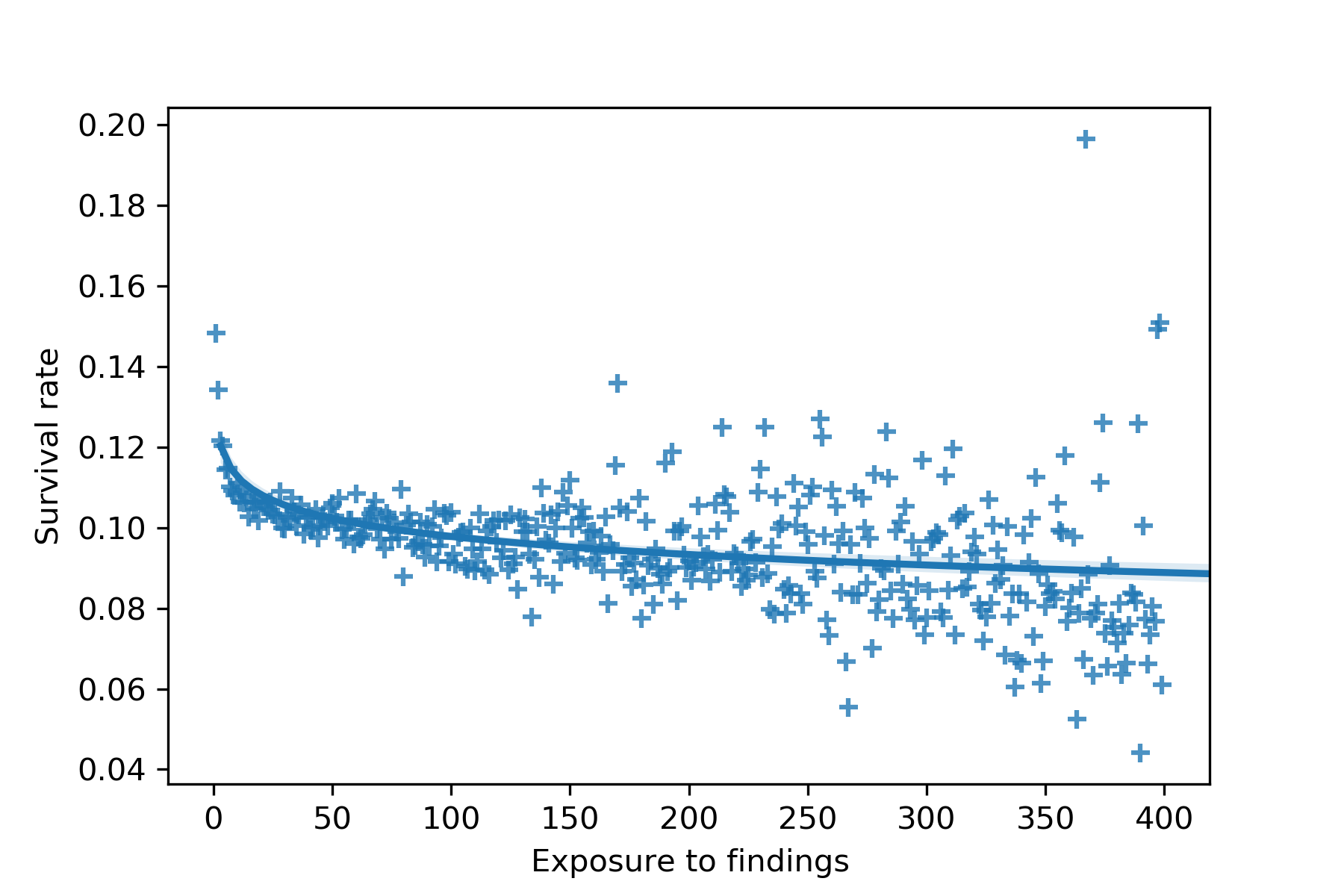}
  \caption {Mutant survival rate as exposure to mutants increases.}
  \label{fig:survival}
\end{figure}

\begin{figure}[t]
  \includegraphics[trim=0 0 0 30, clip, width=\columnwidth]{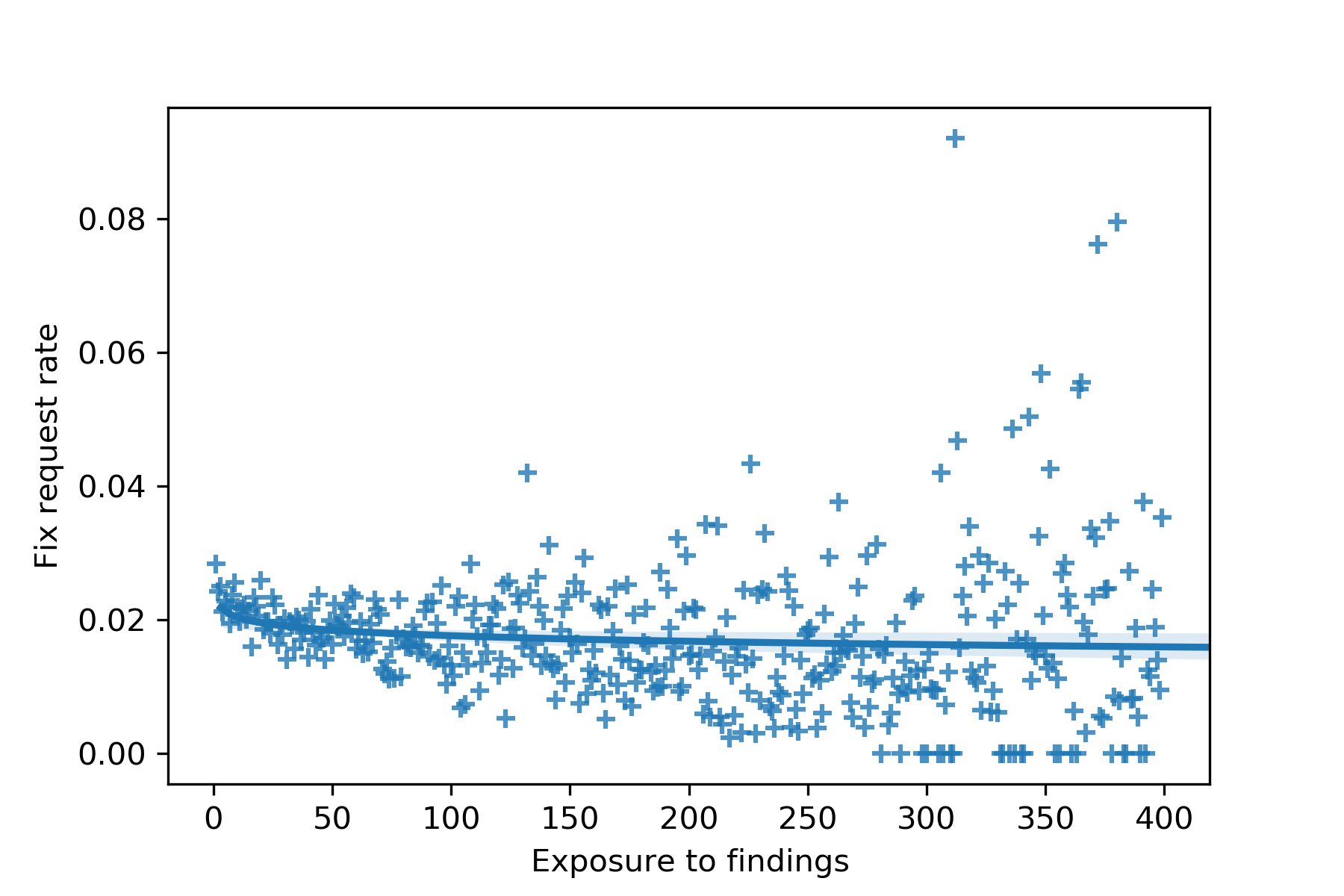}
  \caption {Fix requested rate as exposure to mutants increases.}
  \label{fig:please_fix}
\end{figure}

Figure~\ref{fig:killed} shows the difference between the number of reported live mutants
at the beginning of the code review (\emph{Review}), when code findings are
initially surfaced, and at end of the code review (\emph{Submit}).
Submitted changelists have significantly fewer live mutants, which confirms that
tests added by developers in response to mutants indeed succeed at killing these
mutants.

If additional tests for a changed file improve overall test suite quality, then
future changes to the same file should see fewer surviving mutants.
Figure~\ref{fig:survival} shows the probability of a mutant to survive over time
as files are mutated over and over again: The more mutants a file sees, the more
likely it is that the already existing tests (incl. those written for the
submitted change) kill those mutants. There is a negative relationship
($r_s=-0.50$, $p< .001$) between exposure and survivability of mutants.  These
results suggest that developers not only write more tests when exposed to
mutation testing, the tests they write indeed improve test suite quality.

As a further indication that developers write good tests,
Figure~\ref{fig:please_fix} shows the ratio 
of reported mutants for which the reviewers request developers to write tests.
The results show that, in the beginning, reviewers ask for a larger ratio of the reported
mutants to be killed than later ($r_s=-.34$, $p< .001$). This corroborates that
developers write better tests, such that the surviving mutants
represent less critical issues or futile test goals and reviewers have to
request fewer changes related to reported mutants.
%
%

\begin{result}
  \textbf{RQ2:} As exposure to mutation testing increases, developers tend to
                write stronger tests in response to mutants.
\end{result}

\mypara{Alternative Hypotheses} While we observe that developers write tests
that kill reported mutants, it is theoretically possible that they write minimal
tests that are simple and purely focused on killing mutants, as opposed to tests
they would write otherwise.
However, two general observations and two analyses suggest that this is not the
case.
First, this is unlikely in principle: Code reviews are a safety net and minimal,
change-detector tests would not pass code review.
Second, anecdotally from manually inspecting numerous tests written for mutants
over the past six years, we did not observe any unique features of such tests.
Third, our analysis of mutant redundancy (Section~\ref{sec:majority}) shows that
most mutants generated for the same line are killed by the tests developers
write. Since multiple mutants related to conditional or relational operators
would require distinct tests to be killed (e.g., testing boundary conditions in
a conditional statement), this suggests that developers do not write minimal
tests focused on a single mutant.
Finally, the increasing number of test hunks and the decreasing survivability
(Section~\ref{sec:rq1}) support the conclusion that developers write more tests
than would be necessary to detect individual, reported mutants.  
In summary, we did not find evidence \mbox{that developers write minimal tests for
reported mutants.}

\section{Fault Coupling}
\label{sec:faultcoupling}

The coupling effect is a foundational assumption underlying mutation testing:
Simple faults are coupled to more complex faults in such a way that a test suite
that detects simple faults is sensitive enough to likely detect complex
faults as well. Since mutants represent simpler faults, compared to more complex
real world faults~\cite{GopinathJG2014,JustJE2014}, it is important to assess
whether and to what extent the coupling effect can be observed for mutation
testing of real-world software systems.
In particular, mutants are only suitable test goals that are worth satisfying if
they are coupled to real software faults.

Recall that our mutation testing system only surfaces a single mutant per line
because evaluating all mutants is prohibitively expensive and reporting all live
mutants to a developer in an actionable way is impracticable. However, we
also observed that developers get accustomed to mutation testing in two ways.
First, they use and interpret mutants as concrete guides to improve their code
and corresponding test suites, as opposed to merely writing a minimal test per
test goal (mutant). Second, they write stronger tests in response to, or
anticipation of, mutants that are generated and surfaced.

We hypothesize that most mutants per line share the same fate and that it is
thus sufficient to surface only one of them. Specifically, we hypothesize that if a
developer writes a test suite based on one surfaced mutant in a given line, then
that test suite will detect most of the mutants that could be generated for that
line. Conversely, if a surfaced mutant in a given line survives, then
most mutants that could be generated for that line will survive. Testing
this hypothesis is important to determine whether the assumed
coupling effect holds for mutation testing systems that rely on sampling
representative mutants.


This section answers the remaining two research questions:

\begin{itemize}

\item {\bf RQ3 \Remark{RQ3}}. Are reported mutants coupled with real software
faults? Can tests written based on mutants improve test effectiveness for real
software faults?

\item {\bf RQ4 \Remark{RQ4}}. Are the mutants generated for a given line redundant? Is
  it sufficient in practice to select a single mutant per line?
  
\end{itemize}

\subsection{Dataset}

In order to evaluate whether mutants are coupled to bugs that matter in
practice, we mined our source-code repository and obtained a set of bugs that
had a high priority for being fixed. For each bug, the subsequent coupling
analysis evaluated mutation testing on the buggy and fixed source-code
version of that bug. Specifically, we automated the following analysis:

\begin{enumerate}
  \item Find explicit bug-fixing changes in the version control history
    (Section~\ref{sec:change_selection}).
  \item Analyze changes and retain suitable bugs
    (Section~\ref{sec:change_filter}).
  \item Generate mutants and perform mutation testing on both the buggy and the
    fixed version of the code (Section~\ref{sec:change_mutation}).
  \item Analyze the mutation testing results and identify fault-coupled mutants
    (Section~\ref{sec:change_report}).
\end{enumerate}

\subsubsection{Change Selection} \label{sec:change_selection} We mined our
source code repository for changes that explicitly fix high-priority bugs. Bugs
are assigned a priority class by developers, and high-priority bugs are expected
to be fixed quickly. Our selection approach did not filter for change size or
other code-related attributes.  This was a conscious design decision to obtain
an unbiased sample, spanning multiple programming languages, projects, code
styles, etc.
Our selection step resulted in \textbf{\nBugsTotal candidate bugs}, suitable for our
coupling analysis.

We limited the selection to changes that fixed a bug in the last six months.
This temporal restriction increases the chances of being able to build the
source code and reproduce a bug---the underlying build infrastructure guarantees
that recent source code can be built. Furthermore, we limited the selection to
changes authored by a developer, and hence changes that are subject to code
review. The selection was not limited to changes for which mutation testing was
not enabled. Precisely determining when a bug was introduced and whether
mutation testing was enable for that project at that time is extremely
difficult. To approximate the ratio of bugs that were introduced while
mutation testing could have been enabled, we determined the ratio of bug-fixing
changes for which mutation testing was enabled. Overall, mutation
testing was enabled for 10.8\% of bug-fixing changes in our final dataset.

We rely on developers' change descriptions and
labeling to identify bug fixes.  While bug hygiene is imperfect and not all of
these bugs are necessarily bugs encountered in the field, all of the change descriptions
explicitly claimed to fix a high-priority bug, making it very unlikely that the change
in fact does not.

\subsubsection{Change Filtering}
\label{sec:change_filter}

To establish the coupling between mutants and bug fixes, we need to parse and
build the code, mutate it, and evaluate affected tests against each mutant. This
means that the source code in question needs to be correctly configured,
buildable, and the number of tests to be executed needs to be reasonable.

Mutant generation is non-trivial and carries a hidden cost. For example, our
mutation system uses the \texttt{Clang} compiler front end for the AST analysis and manipulation of C++ code.
This requires indexing the compilation
unit with the build system to prepare the transitive closure of included headers
and the set of \texttt{Clang} arguments that are mandated by the project-under-mutation's
configuration. A common example is code that runs on specialized architecture,
or requires additional virtualization libraries, which is unknowable given just
the file, and the file is not analyzable without it. The additional cost of
indexing all C++ compilation units is relatively small, but still
accounts for a large resource expenditure overall.

While the total number of generated mutants is comparatively low due to our
mutant suppression and reduction strategies, the sheer number of required test
target executions renders this coupling analysis as very expensive.  A
\emph{test target} defines a set of tests at the build system level. A test
target can be a single test file or a test suite (i.e., collection of test
files)---each potentially containing hundreds or thousands of individual tests.
This variance comes from how the build system is used, and differs from project
to project.
In a large project with complex dependencies, any change to a widely used core
library will easily require millions of tests to be executed, each of which has
the said library in the transitive closure of its dependencies, and all of which
have to be evaluated. Failure to evaluate all test targets, as is the norm,
inevitably leads to inaccuracies of reported live mutants and would decrease
developers' confidence in mutation testing.

\begin{figure}
  \includegraphics[width=\columnwidth]{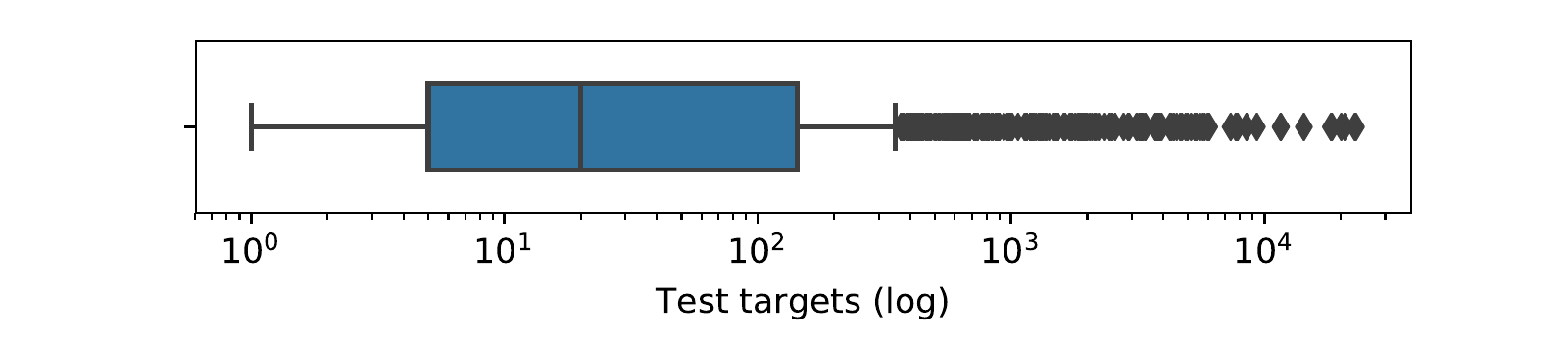}
  \caption {Number of test targets for all candidate bugs.}
  \label{fig:test_count_all}
  \includegraphics[width=\columnwidth]{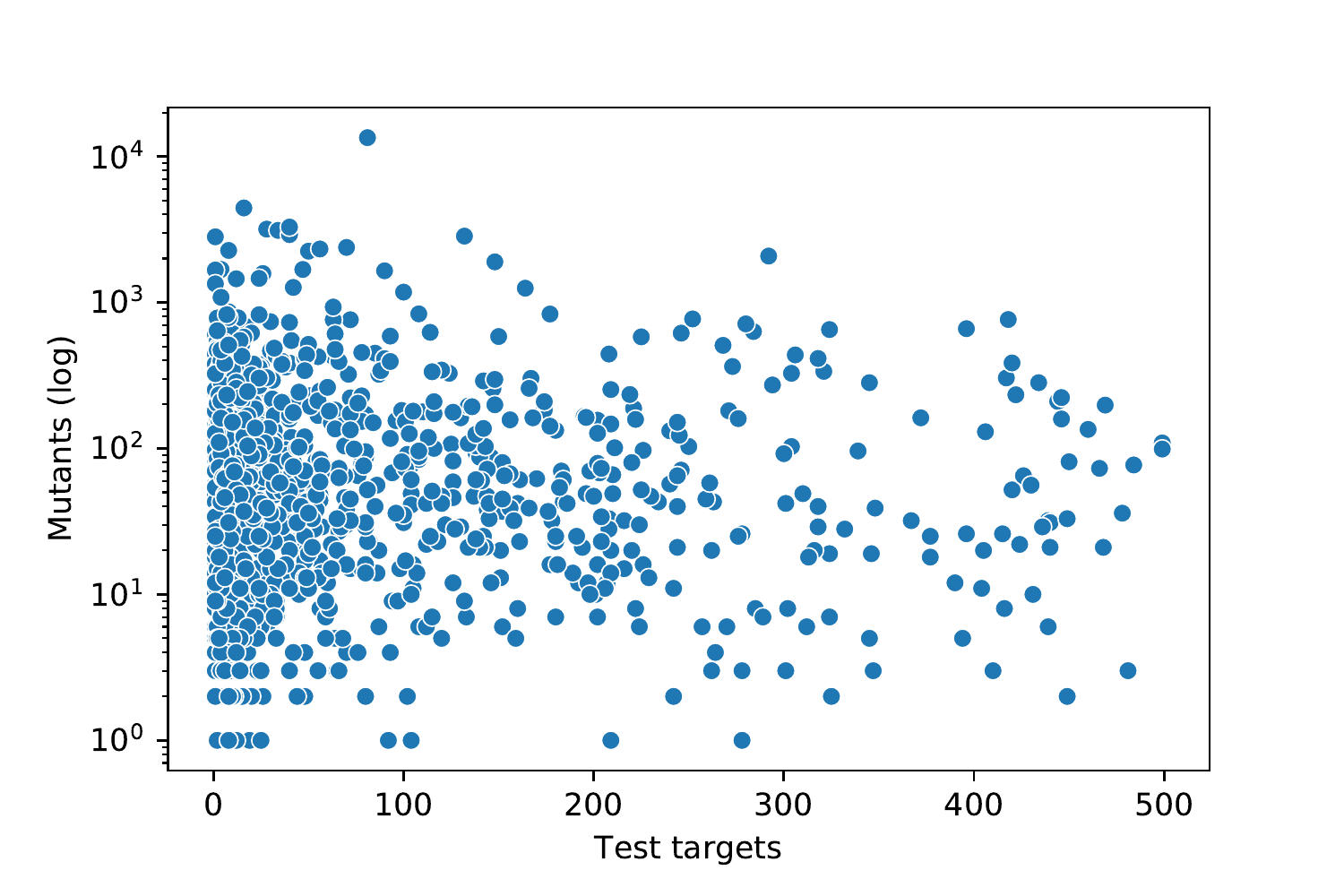}
  \caption{Number of test targets and number of mutants for all retained bugs.}
  \label{fig:test_count}
\end{figure}

Given the challenges above and our \nBugsTotal candidate bugs, we automatically
filtered:
\begin{itemize}

\item \nBugsNonRepro bugs for which a project configuration failure prevented us from
building the buggy or fixed source code version, rendering them unsuitable for
our analysis.

\item \nBugsTooLarge bug fixes for which the project configuration exhibited a substantial
dependency chain with a very large number of associated test targets. Including these
bugs would have made our analysis computationally infeasible. We empirically
determined a threshold of 500 test targets (each usually containing many
tests) as our filter criterion. Figure~\ref{fig:test_count_all} shows the
distribution of the number of test targets for all \nBugsTotal candidate
bugs---indicating that bugs with more than 350 test targets are outliers.

\end{itemize}

Our final dataset contains \textbf{\nBugsFinal retained bugs}.
Table~\ref{tab:bugs} shows further characteristics of these bugs,
and Figure~\ref{fig:test_count} shows their distribution of number of
test targets and number of mutants. In total, our coupling analysis
involves \nBugsFinal bugs, almost 400 thousand mutants, and over 33 million test target
executions.
We performed the coupling analysis over a prolonged period of time because of the huge
amount of processing power required for the dependency analysis,
mutagenesis, and test target executions.

\begin{table}
  \caption{Characteristics of the \nBugsFinal retained bugs.}
  \label{tab:bugs}
  \setlength{\tabcolsep}{3pt}
  \resizebox{\linewidth}{!}{%
  \begin{tabular}{l rr rr rr rr}
    \toprule
     & \multicolumn{2}{c}{\textsc{Bugs}} &
                        \multicolumn{2}{c}{\textsc{Test targets}} &
                        \multicolumn{2}{c}{\textsc{Affected files}} &
                        \multicolumn{2}{c}{\textsc{Affected lines}}\\
    \cmidrule(lr){2-3}
    \cmidrule(lr){4-5}
    \cmidrule(lr){6-7}
    \cmidrule(lr){8-9}
    & \textsc{Count} & \textsc{Ratio} &
      \textsc{Median} & \textsc{Mean} &
      \textsc{Median} & \textsc{Mean} &
      \textsc{Median} & \textsc{Mean} \\
    \midrule
    C++ & 736 &  49.0\% & 12 & 35.0 & 3 & 4.8 & 52 & 174.7\tabularnewline
Java & 501 &  33.4\% & 22 & 42.5 & 3 & 4.9 & 50 & 263.3\tabularnewline
Python & 180 &  12.0\% & 10 & 30.7 & 2 & 3.5 & 24 &  83.6\tabularnewline
Go & 85 &   5.7\% & 5 & 22.2 & 3 & 3.6 & 41 & 516.9\tabularnewline
\midrule
Total & 1502 & 100.0\% & 13 & 36.2 & 3 & 4.6 & 46.5 & 212.7\tabularnewline

    \bottomrule
  \end{tabular}%
}
\end{table}

\subsection{Methodology}
\label{sec:change_mutation}

Incremental mutation testing is substantially different from the traditional
approach. Lines are selectively mutated based on the structure of
the code, mutator operators are probabilistically picked and tried out, and most
potential mutants are discarded because reporting too many mutants during
code review would be too visually daunting and likely have a
negative effect on developers' perception of mutation testing.

In order to evaluate whether the mutants our mutation system generates are
coupled with the \nBugsFinal high-priority bugs that we retained, we had to modify
the system to generate multiple mutants per line and to evaluate all of
them. This turned out to be very expensive even for a manageable number of bugs,
validating our intuition that the traditional approach would not scale to an
environment with thousands of changes per day. Note that we kept the notion of
suppressing unproductive mutants~\cite{PetrovicIKAJ2018}, which represent futile
test goals, in the modified system, and hence did neither generate nor evaluate
them.

For each bug in our dataset, we automatically executed the following
steps:

\begin{enumerate}
\item Run dependency analysis and determine all test targets affected by the bug-fixing
      change.
\item Using the buggy source code version, generate all possible mutants for all
      lines that are affected by the bug-fixing change.
\item Using the fixed source code version, generate all possible mutants for all
      lines that are affected by the bug-fixing change.
\item Determine the set of fault-coupled mutants.
\end{enumerate}

In line with prior work~\cite{JustJIEHF2014b}, we measure fault-coupling as
follows: Given a test suite with at least one fault-triggering test, a buggy
source code version, and a fixed source code version, a mutant is coupled to a
fault (through the triggering tests) if (1) that mutant exists in the buggy and
the fixed source code version and (2) that mutant is only detected by the
triggering tests. In other words, a fault-coupled mutant is live in the buggy
but killed in the fixed source code version: Had mutation testing been run on the buggy source code
version, a mutant would have been reported in the lines affected by the fixing
change, and the test accompanying the fix kills it.

\subsection{RQ3: \Remark{RQ3}}
\label{sec:change_report}

This research question is concerned with whether mutation testing would have
reported a live mutant on the change introducing the bug---a live mutant that
is subsequently killed by a bug-triggering test.  Reporting that mutant would
have had a good chance of preventing the bug being introduced in the first place.

\subsubsection{Coupled Faults}

\begin{figure}
\centering
\includegraphics[width=.95\linewidth]{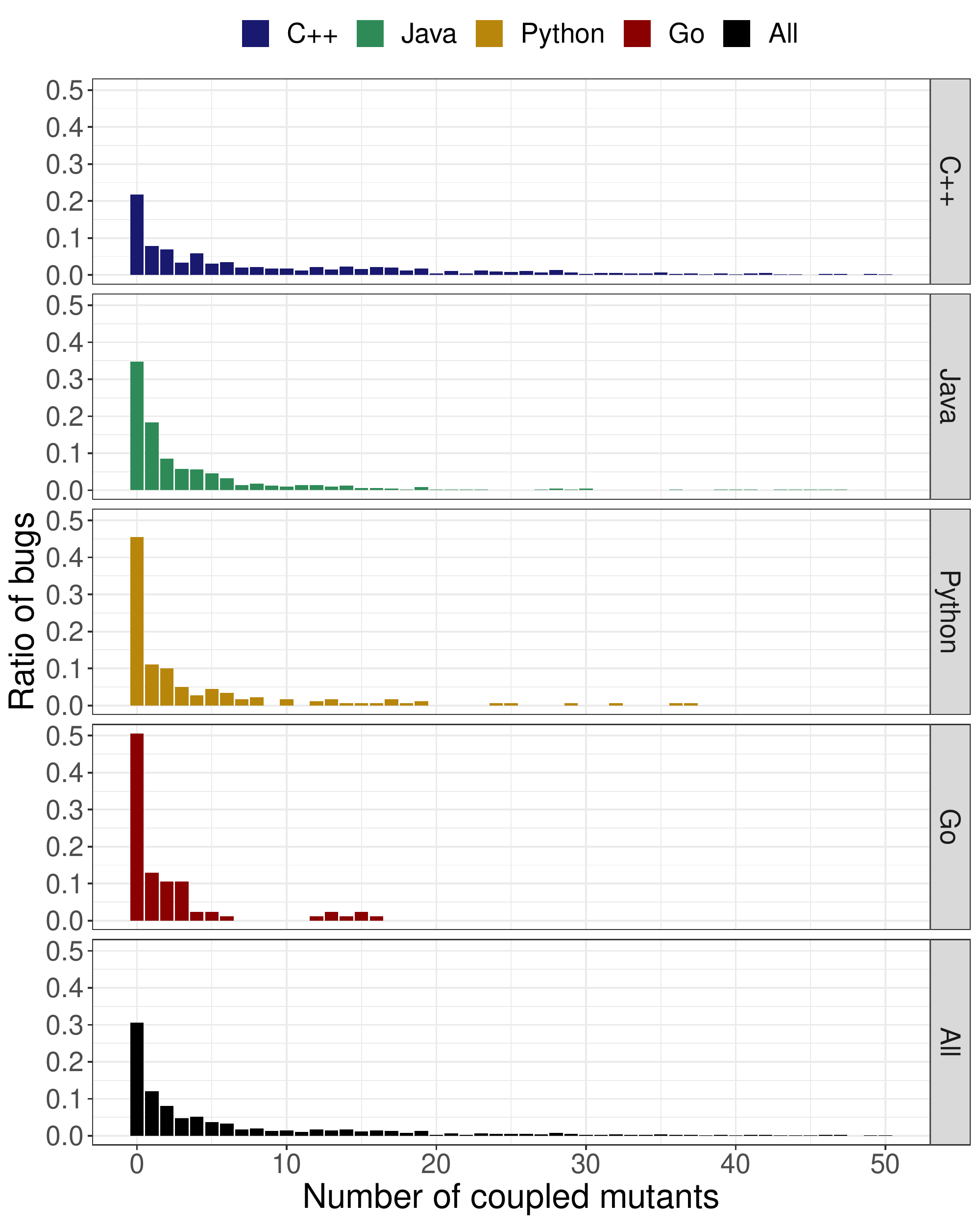}
\caption{Number of mutants coupled to each of the \nBugsFinal bugs, grouped by
programming language of the bug fix. 4\% of all bugs
(51 to 485 coupled mutants) are ommitted for clarity.\label{fig:coupling-bugs}}
\end{figure}

We found that for \nBugsCoupled (\ratioBugsCoupled) of the bugs, mutation
testing would have reported a fault-coupled mutant in the bug-introducing
change. Recall that each bug-introducing change was covered by the existing
tests, suggesting that code coverage had exhausted its usefulness.

\begin{result}
  \textbf{RQ3:} Mutants are coupled with \ratioBugsCoupled of high-priority
                bugs, for which mutation testing would have reported a live,
                fault-coupled mutant on the bug-introducing change.
\end{result}

It is interesting to note that when a bug is coupled to a mutant, it is
usually coupled to more than one, as seen in Figure~\ref{fig:coupling-bugs}.
This observation is consistent with the finding that the majority of mutants, generated for a given line, share the same fate (Section~\ref{sec:majority}).

\begin{figure}
\centering
\includegraphics[width=.95\linewidth]{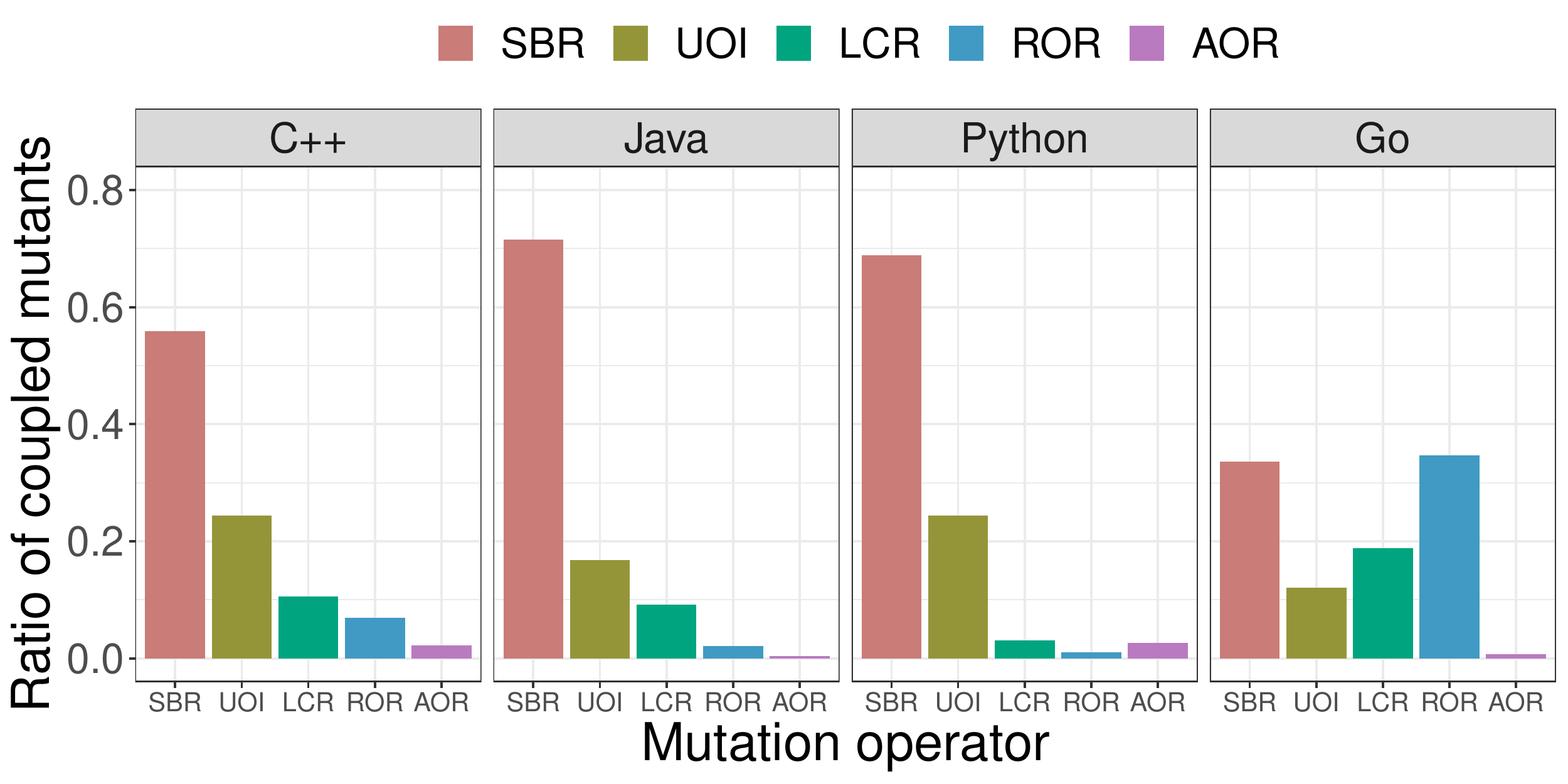}
\caption{Ratio of coupled mutants per mutation operator, grouped by
programming language of the bug fix.\label{fig:coupling-ops}}
\end{figure}

The statement block removal (SBR) mutation operator, being the most prolific one,
generated most of the coupled mutants (Figure~\ref{fig:coupling-ops}).
The distributions of number of coupled mutants
are consistent across all languages (Figure~\ref{fig:coupling-bugs}).
In contrast, the distribution of coupled mutants across mutation operators is
different for Go (Figure~\ref{fig:coupling-ops}). Given the relatively small
sample size of Go bugs, it is possible that this observation is an artifact of
the sample size or that Go bugs have indeed different coupling characteristics.

\subsubsection{Non-coupled Faults}

Recall that \nBugsNotCoupled out of \nBugsFinal (\ratioBugsNotCoupled) bugs were not coupled to
any of the generated mutants. We randomly sampled 50 of the non-coupled bugs for
manual inspection. For each bug, we determined the reason for the absence of
fault coupling, following the classification of Just
\etal~\cite{JustJIEHF2014b}: weak mutation operator, missing mutation
operator, no such mutation operator.

\mypara{Weak mutation operator (14/50)}
While focusing our analysis on the weaknesses of the four mutation
operators employed by our mutation system, two patterns emerged. First, the SBR
mutation operator currently does not
mutate statements that affect control flow, including return, continue,
and break statements. The reason is that mutating these statements is more
likely to cause compilation issues or infinite loops. Given the occurrences of
these statements in our bug data set, however, it seems worthwhile to strengthen the SBR mutation operator
to mutate them, employing additional heuristics or program analyses to avoid
invalid mutants. 
Second, the SBR mutation operator does not delete variable declarations because
these mutants would not compile. However, a variable declaration with a
(complex) initializer should be mutated, e.g., by replacing the initializer with
a constant value.

\mypara{Missing mutation operator (13/50)}
Recall that our set of mutation operators and the corresponding suppression
rules are intentionally chosen to make mutation testing scalable and the number
of surfaced mutants manageable. Nonetheless, we considered a broad set of
mutation operators, regardless of their costs, to determine whether additional
mutation operators would generate mutants that are coupled to the bugs in our
data set. We noticed that identifier-based bugs~\cite{PradelS2018} are a nontrivial portion in our
randomly selected set of non-coupled bugs. While prior work demonstrated that
mutation operators that target identifiers have the potential to increase fault
coupling~\cite{AllamanisBJS2016,WildCaught}, the same work also showed that these mutation
operators can easily quadruple the number of mutants. We leave a deeper
investigation into heuristics and program analyses to tame these mutation
operators for future work.

\mypara{No such mutation operator (23/50)}
Similar to the findings of Just \etal~\cite{JustJIEHF2014b}, we observed a
number of non-coupled bugs for which no obvious mutation operator exists.
Examples include very subtle, yet valid changes to configurations or
environments, for which general-purpose mutation operators are not applicable.
Similarly, bugs in higher-level specifications or protocols are outside of the
scope of mutation testing. In other words, mutation testing is effective in
guiding testing to assess whether an algorithm is correctly implemented but not
whether the correct algorithm is implemented.

\subsubsection{Discussion}

While we desire higher fault coupling (to prevent more bugs), we also have a
competing incentive to show only productive mutants and not overwhelm the
developers.  Generating more mutants would almost certainly increase fault
coupling, but at what cost? If 1\% more coupling means 100\% more mutants, or
worse 100\% more unproductive mutants, then this is undesirable:
Developers would likely abandon mutation testing because of too many ``false
positives''.  It is possible that fault coupling can be increased without
increasing the ratio of unproductive mutants by adding additional suppression
rules tailored to additional mutation operators.

Fault coupling is a valid measure if tests written for mutants are similarly
effective as those that are written for other objectives.
Section~\ref{sec:longtermeffects} provided evidence that this is indeed the
case. Further, we count on code authors to push back on introducing tests to
kill unproductive mutants, and we count on reviewers to push back on low quality
tests, specifically written to kill mutants. Anecdotally, we often
see both code authors and reviewers pushing back on a mutant because the kind of
test that would kill it is considered low quality or even harmful. Over the past six
years, we have observed that code authors do not blindly use mutants as test
goals. Rather, they reason about their usefulness and report unproductive
mutants so that we can improve suppression rules.  Overall, we have no reason to
believe that tests written for the mutants are substantially different from
other tests.

\subsection{RQ4: \Remark{RQ4}}
\label{sec:majority}

Our mutation testing system rests on the assumption
that generating and evaluating multiple mutants in the same line is not
necessary. The primary reason for this is that we do not compute the mutation
score but rather report mutants as test goals and direct developers' attention
to the mutated piece of code---any one \emph{productive}
mutant is sufficient for the latter. A secondary reason is the
expectation that mutants for a given line are highly redundant.

We tested our hypothesis about mutant redundancy by analyzing the mutant data
from our data set of retained bugs. Since we generated all possible mutants for that data
set, we can reason about redundancy.  Specifically, we looked at the testing
outcomes of mutants in lines for which multiple mutants were generated. We
calculated the ratio of the majority event for each line---that is, the ratio of
mutants in the majority group sharing the same fate (live or killed).  This
allows us to capture both cases when most mutants are killed and when most
mutants are not; the closer the majority fate is to 100\%, the higher the
redundancy.
The rationale for validating our hypothesis is twofold. First, the results
can inform mutant sampling strategies. Second, the results from our coupling
analysis translate to our own mutation system only if the majority fate is high
on average---that is, if picking a single mutant per line is indeed sufficient.

\begin{figure}
  \includegraphics[page=1,width=.95\linewidth]{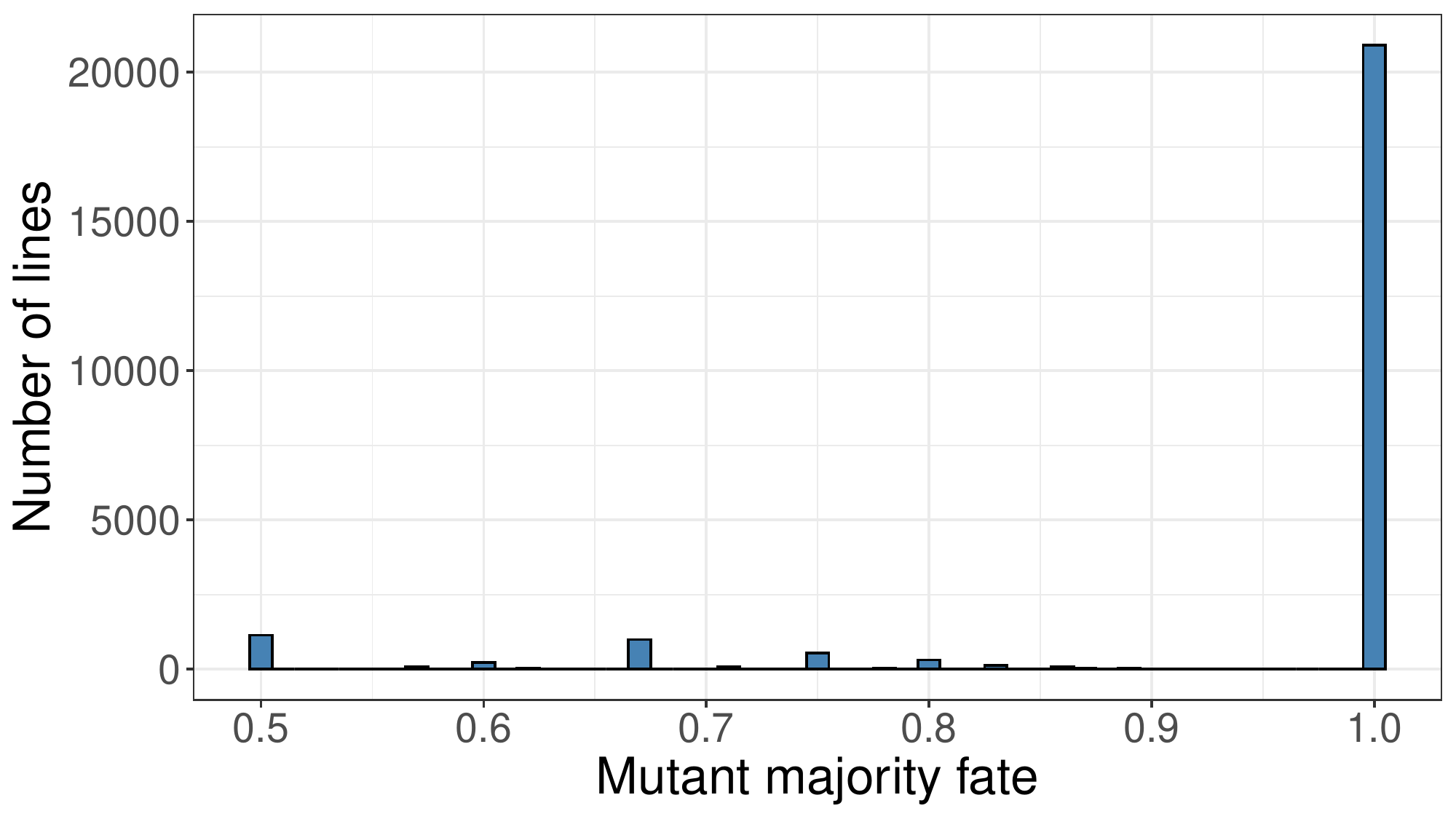}
  \caption {Mutant majority fate for all lines with multiple mutants.}
  \label{fig:majority_fate}
\end{figure}

Figure~\ref{fig:majority_fate} shows the distribution of the mutant majority
fate for all lines that have at least to mutants.
Given that more than 90\% of all lines have a mutant majority fate of 100\%,
we conclude that the generated mutants are highly
redundant and that generating and reporting at most a single mutant per line is
a valid optimization.

\begin{result}
  \textbf{RQ4:} Mutants are heavily redundant. In more than 90\% of cases,
  either all mutants in a line are killed, or none are.
\end{result}

\section{Threats To Validity}
\label{sec:threats}

\textit{Construct validity} is concerned with the chosen proxy measures and whether these
accurately measure the concept of interest. To address this concern, we relied
on proxy measures accepted in the literature, where applicable. Given that there
are no comparable studies on the long-term effects of mutation testing, we had
to make two notable choices. First, we measure exposure to mutants as the number
of times mutants are surfaced during code review, on a per-file basis. We argue
that this is a valid proxy for exposure because (1) both authors and reviewers
have exposure to mutants during the code review process and (2) code review
usually involves peer review of project members, and hence this measure captures
the exposure of a development team to mutation testing.
Second, we measure testing effort as the number of changed test hunks (as opposed
to the raw lines of code). This was a concious choice because lines of code do
not generalize across programming languages or testing paradigms.
Finally, we consistently employed baselines for comparisons, and the results not
only show a signal for mutation testing but also the absence of a signal for the
coverage-testing baseline.

\textit{Internal validity} is concerned with how well our study design isolates
variables of interest and whether it accounts for possible confounding.  To
address this concern, we relied on an interventional study design and tested
alternative hypotheses for all
reported observations, which our study suggests are linked to mutation testing,
thereby increasing confidence in our results. For example, when quantifying how
exposure to mutants affects testing effort, we considered possible confounding
effects, such as number of affected
files and an increase in code coverage. Our explorations of alternative
hypotheses all led to the same conclusions. It is, however, possible that we
missed other confounding effects.

\textit{External validity} is largely concerned with generalizability and how
well the reported results translate to other development environments.  Our
study reports on data and observations from a single company. However, our study
involves tens of thousands of developers and many different projects, which
we believe are representative of a larger population. Furthermore, contemporary
code review is ubiquitous and used by software engineers at other
companies and in open-source projects. 

\section{Related Work}
\label{sec:related}

While mutation testing has seen growing interest in research and practice,
reports of actual deployments and studies of its efficacy and long-term effects
are very rare in the literature. Indeed, we are not aware of any studies that
investigate the long-term effects of mutation testing on test quality, test
quantity, and developer behavior.
This section describes the prior work most closely related to ours, focusing on
industrial case studies, studies that compare the characteristics of mutants and
real faults, and existing empirical evidence that suggests that mutation testing
is likely effective in practice.

\mypara{Industrial case studies}
Ahmed \etal reported on a case study, performing mutation analysis with 3169
mutants for a Linux Kernel module, with subsequent analysis of 380 surviving
mutants~\cite{AhmedKernelTestSuites}.  The study aimed at reducing the
computational costs of performing mutation analysis on complex software systems,
and it concluded that ``mutation testing can and should be more extensively used
in practice''.

Delgado-P{\'e}rez \etal reported on a case study, performing mutation analysis
with 2509 mutants for 15 functions (ranging from 10 to 63 lines of code) of
different firmware modules~\cite{Delgado2018}, with subsequent manual analysis
of 154 surviving mutants. This study focused on the computational costs and
human effort for identifying equivalent mutants and developing a
mutation-adequate test suite by extending a coverage-adequate one. The study
concluded that ``mutation testing can potentially improve fault detection
compared to structural-coverage-guided testing''.

In prior work, we reported on the scalable
mutation testing system deployed at Google~\cite{PetrovicI2018},
as well as on challenges associated
with applying mutation testing in practice~\cite{PetrovicIKAJ2018}.
In this prior work, we addressed the
computational costs of applying mutation testing at scale, with a key focus on
the identification and elimination of unproductive mutants---mutants that
developers consider non-actionable test goals (similar to false positive
warnings in static analysis). In other words, we identified mutants that can but
should not, and in practice will not, be detected to avoid ineffective tests
that negatively affect testing time and maintainability.

In contrast to the three industrial case studies above, our work differs in two
ways. First, it reports on a longitudinal, interventional study that spans six
years of development and involves more than 14 million mutants, reporting on the
effects of mutation testing on test quantity and quality.
Second, it reports on whether critical real faults are coupled to
mutants---whether \mbox{mutation testing has the potential to prevent those
faults.}

Very recently, an industrial application of mutation testing at
Facebook~\cite{beller2021use} applied complex mutation operators learned from
past bug-inducing changes. These operators achieve higher survival rates at the
cost of being applicable to smaller parts of the codebase. These learned
mutation operators could offer an alternative approach to taming the number of
mutants. Similar to our mutation testing system, Facebook's
system provides information to developers during code review.

\mypara{Fault coupling and fault characteristics}
A number of empirical studies showed that mutants are coupled to real faults and
that mutant detection is positively correlated with real fault
detection~\cite{DaranT1996,AndrewsBL2005,AndrewsBLN2006,JustJIEHF2014b,ChenGTEHFAJ2020}. The same studies
also showed limitations and that about 27\% of real faults were not coupled with
commonly generated mutants. Brown \etal~\cite{WildCaughtMutants2017} and
Allamanis \etal~\cite{AllamanisBJS2016} aimed at narrowing this gap with their
work on wild-caught mutants and tailored mutants, respectively. Both approaches
employ semantics-related mutations (e.g., replacing identifiers with similar,
type-compatible alternatives) and have the potential to further improve fault
coupling. However, the improved fault coupling comes at a cost of significantly
more mutants. For example, replacing function calls with all type-compatible
alternatives more than quadrupled the number of mutants.
Gopinath \etal analyzed the relationship between mutants and real faults from a
different viewpoint. Specifically, they compared the complexity and
distributions of mutants and real faults~\cite{GopinathJG2014}. Their analyses
showed that a typical real fault is more complex in terms of syntactical tokens
and that real faults are rarely equivalent to mutants generated by traditional
mutation operators.

The observed positive correlations between mutant detection and real fault
detection on the one hand and the different characteristics of mutants and real
faults on the other hand, motivated, in part, our work on studying the long-term
effects of mutation testing and its efficacy in practice.

\mypara{Mutant characteristics}
Other researchers have addressed the notion that some mutants are more valuable
than others, including
{stubborn} mutants (Yao \etal~\cite{YaoStubbornICSE14}),
{difficult-to-kill} mutants (Namin \etal~\cite{NaminMuRanker2014}),
{dominator} mutants (Kurtz \etal~\cite{KurtzMSGsICSTW2014} and Ammann \etal~\cite{MinimalMutantsICST2014}),
and
{surface} mutants (Gopinath \etal~\cite{GopinathMutation2016}).
While these definitions are useful in a research context (e.g., to study
redundancy among mutants), they are not directly relevant to a developer in
practice. For example, a difficult-to-kill mutant may still be
unproductive~\cite{PetrovicI2018}, and hence writing a test for it would be undesirable.

\mypara{Mutant sampling}
Prior work examined whether guided mutant sampling is more effective
than random mutant sampling for similar numbers of mutants.  Acree \cite{AcreeDiss80}
and Budd~\cite{BuddDiss80} independently concluded that a test suite developed to
detect a randomly selected 10\% of mutants is almost as effective as a test
suite that detects all of the mutants.
Wong and Mathur~\cite{WongReducingCost1995} reached similar conclusions, finding
that randomly sampling mutants beyond 10\% yields marginal improvements.
Zhang \etal~\cite{ZhangOperatorVsRandom2010} compared guided mutant sampling
and random mutant sampling and also found no appreciable difference in their
performance.
Gopinath \etal~\cite{Gopinath2015} expanded this investigation, using a large
body of open-source software, again finding that random mutant sampling
performs as well as any of the considered guided mutant sampling strategies.

In line with prior results, our mutation approach samples very few mutants to
make mutation testing applicable at scale. Our results confirm that the set of
mutants, generated with traditional mutation operators, is indeed highly
redundant and that small sampling ratios are sufficient. Moreover, our results
show that multiple mutants generated for the same line of code virtually always
share the same fate in practice, implying that surfacing one mutant per line is
sufficient. This observation is consistent with Zhu et al.'s work~\cite{Zhu2018}
that also supports the selection of representative mutants.

\section{Conclusions}

The idea of mutation testing was introduced more than four decades ago, and in
all this time research revolved mainly around
problems of scalability. All along, research was based on the fundamental
assumption that mutants are meaningful and actionable test goals which lead to
positive effects. This assumption, however, has not been evaluated until now.

We have implemented and deployed a mutation testing system capable of scaling to
very large software. Developers are shown only a fraction of the possible
mutants, by limiting the number of mutants and suppressing mutants assumed to
be unproductive test goals.
This paper uses long-running production data of this system in order to
validate the central assumption underlying mutation testing for the very first
time.

Our results show that developers working on projects with mutation testing
write more tests on average over longer periods of time, compared to
projects that only consider code coverage. Mutants are effective test
goals: Developers exposed to mutants write more, effective tests in
response to them.

Since the ultimate goal is not just to write tests for mutants, but to prevent real bugs,
we investigated a dataset of high-priority bugs and analyzed mutants before and after
the fix with an experimental rig of our mutation testing system. In 70\% of
cases, a bug is coupled with a mutant that, had it been reported during code
review, could have prevented the introduction of that bug.
Finally, we also validated our approach of generating a single mutant per line:
For the vast majority of lines, either all mutants in a line are killed, or
all survive. Therefore, generating a single mutant per line is a valid
optimization.

This paper finally provides evidence that mutants are indeed
meaningful and actionable test goals. Considering that this insight emerges at a
time where robust, industry-strength mutation tools appear for more and more
programming languages, we hope that mutation testing will see a substantial
boost in industrial adoption, leading to better software quality.

\section*{Acknowledgements}
This work is supported in part by EPSRC project EP/N023978/2 and by
the National Science Foundation under grant CCF-1942055.

\balance
\bibliographystyle{IEEEtran}
\bibliography{IEEEabrv,bib/bib-constants,bib/google,bib/mutation,bib/coverage,bib/rjust,bib/other}

\end{document}